\documentclass[10pt,a4paper,aps,prl,reprint,nobibnotes,floatfix,superscriptaddress,showpacs]{revtex4-1}

\usepackage[utf8]{inputenc}
\usepackage[english]{babel}

\usepackage{amsfonts}
\usepackage{amsmath}
\usepackage{amssymb}
\usepackage{bbold}
\usepackage[dvipsnames,svgnames,table]{xcolor}
\usepackage{graphicx}
\usepackage{fancyhdr}
\usepackage{float}
\usepackage{braket}
\usepackage{float}
\usepackage[caption=false]{subfig}
\usepackage{tikz}  
\usepackage{chngcntr}
\usepackage{ulem}

\usepackage[dvipsnames]{xcolor}

\usepackage{hyperref}
\hypersetup{
    pdfstartview={FitH},
    colorlinks=true,    
    linkcolor=NavyBlue, 
    citecolor=Maroon,   
    filecolor=NavyBlue, 
    urlcolor=NavyBlue   
}

\makeatletter
\def\footnoterule{\kern -10pt
    \hrule \@width 100pt \kern 10pt} 
\makeatother

\begin{document}


\title[]{Chiral spin-transfer torque induced by curvature gradient}

\author{G. H. R. Bittencourt}
 \affiliation{Universidade Federal de Vi\c cosa, Departamento de F\'isica, \\Avenida Peter Henry Rolfs s/n, 36570-000, Vi\c cosa, MG, Brasil.
 }
 
\author{M. Castro}
 \affiliation{
 Universidad de Santiago de Chile, Departamento de F\'isica, Cedenna, Avda. {V\'ictor Jara} 3493, Estaci\'on Central, Santiago, Chile
 }

\author{A. S. Nunez}
 \affiliation{Departamento de F\'isica, FCFM, Universidad de Chile, Santiago, Chile}
 
 \author{D. Altbir}
 \affiliation{Universidad Diego Portales, Ej\'ercito 441, CEDENNA, Santiago, Chile}

\author{S. Allende}
\affiliation{
 Universidad de Santiago de Chile, Departamento de F\'isica, Cedenna, Avda. {V\'ictor Jara} 3493, Estaci\'on Central, Santiago, Chile
 }

\author{V. L. Carvalho-Santos}
\email{vagson.santos@ufv.br}
\affiliation{%
Universidade Federal de Vi\c cosa, Departamento de F\'isica, \\Avenida Peter Henry Rolfs s/n, 36570-000, Vi\c cosa, MG, Brasil. 
}%


\date{\today}

\begin{abstract}
This work analyzes the propagation of a transverse domain wall (DW) motion under the action of an electric current along a nanowire (NW) with a curvature gradient. Our results evidence that the curvature gradient induces a chiral spin-transfer torque (CSTT) whose effect on the DW motion depends on the direction along which the DW points. The origin of the CSTT is explained in terms of a position and phase-dependent effective field associated with the DW profile and the electric current direction. Finally, our results reveal that this chiral mechanism can also affect the behavior of other magnetization collective modes, such as spin waves. This work shows the emergence of curvature-induced chiral spin transport and highlights a new phenomenon to be considered for designing spintronic devices.
\end{abstract}

\maketitle

The possibility of manipulating the fundamental properties of an electron beyond its charge has given rise to the field of spintronics \cite{Hirohata-JMMM,Hrcak}. This fundamental and applied research field arrives with the promise of developing a new generation of devices based on spin transport. Some of the proposals that use the spin as an information carrier are logic gates \cite{Vander-JPD,Goolap-SciRep}, neuromorphic computing \cite{Torrejon-Nat,Grolier-Nat}, and racetrack memory devices \cite{Parkin-Nat,Parkin-Nat2}. In the latter, information is encoded in magnetic domains that propagate through the motion of magnetic textures such as domain walls (DWs) \cite{Goolap-SciRep,Parkin-Nat,Parkin-Nat2,Indian-IEEE,Tomasello-JPD,Indian-PREP}, skyrmions \cite{Tomasello-JPD,Tomasello-SciRep}, and vortices \cite{Geng-JMMM}. Although multiple studies addressed the propagation of magnetic quasiparticles, racetrack memory faces problems that should be overcome before its technological implementation \cite{Indian-IEEE}. Some difficulties in controlling the motion of magnetic textures and their positions are the existence of a Walker breakdown \cite{Walker,Donahue,Mougin-EPL,Thiavile-EPL}, the Joule heating \cite{Raposo-APL}, and pinning resulting from the presence of defects \cite{Gao-AIP,Yuan-PRB,Djuhana-JMMM} and curvature gradients \cite{Yershov-PRB,Bittencourt-PRB,Santos-APL,Korniienko-PRB}. 

A particular issue regarding the implementation of racetrack memory devices is the occurrence of potential wells when a DW moves along a nanostripe with defects. In this case, the DW experiences pinning effects when it displaces in staggered \cite{Stag-1,Stag-2}, notched \cite{Indian-IEEE,Hayashi-PRL,Hayashi-NatP}, and bent \cite{Yershov-PRB,Bittencourt-PRB,Wang-JJAP,Omari-APL,Parkin-Nat3,Meier-JAP,Lewis-APL,Beach-Nat} NWs. The DW pinning and unpinning effects depend on geometry-induced interactions. One example is the magnetochiral coupling of a DW at the maximum curvature of an elliptical nanowire, where the DW type determines the direction in which its center points \cite{Yershov-PRB,Bittencourt-PRB}. Also, the threshold value of current densities and magnetic fields used to unpin DWs from a bent defect depends on the direction the DW center points and the underlying geometry of the magnetic system \cite{Hayashi-PRL,Wang-JJAP,Parkin-Nat3,Lewis-APL,Beach-Nat}. Although some authors attributed this asymmetric pinning to the dipolar interaction \cite{Wang-JJAP} or to edge roughness in the curved portion of the wire \cite{Parkin-Nat3}, the emergence of exchange-driven geometry-induced effective interactions \cite{Sheka-JPA,Gaididei-PRL} is also responsible for several features regarding these phenomena. These examples indicate that a profound analysis of several aspects behind the magnetization dynamics is fundamental to their well-controlled and reliable propagation along a magnetic body. 

In this letter, through analytical and numerical calculations alongside micromagnetic simulations, we describe the phenomenon here-called curvature-induced chiral spin-transfer torque (CSTT), responsible for yielding a handedness in the DW dynamics when it displaces along a NW with a curvature gradient. Such an curvature-induced chiral spin transport depends on  the DW type and direction it points (DW phase), yielding direction-dependent dynamical behaviors.  

Our system consists of a ferromagnetic NW parameterized as $\mathbf{r}(s)=\boldsymbol{\chi}(s)+\xi_2\,\hat{\mathbf{e}}_2+\xi_3\,\hat{\mathbf{e}}_3$, where $\hat{\mathbf{e}}_2$ and $\hat{\mathbf{e}}_3$ are unitary vectors in a Frenet-Serret basis corresponding to the normal and binormal directions, respectively. Here, $\boldsymbol{\chi}(s)$ determines the NW central line lying within the $xy$ plane, and $s$ is the natural parameter (arc length). The tangential direction is described by $\hat{\mathbf{e}}_1=\boldsymbol{\chi}'(s)$. Additionally, $\hat{\mathbf{e}}_2=\boldsymbol{\chi}''(s)/\kappa(s)$ and $\hat{\mathbf{e}}_3=\hat{\mathbf{e}}_1\times\hat{\mathbf{e}}_2$, being $\kappa(s)\equiv\kappa=|\boldsymbol{\chi}''(s)|$ the NW curvature. The NW cross-section is defined by the parameters $\xi_2 \in [-w/2, w/2]$ and $\xi_3\in [-h/2, h/2]$, where $w$ and $h$ are the cross-section width and height, respectively. The magnetization can be parametrized using the curvilinear basis as $\mathbf{m} = \cos\Omega\, \hat{\mathbf{e}}_1 + \sin\Omega\cos\Phi\, \hat{\mathbf{e}}_2 + \sin\Omega\sin\Phi\, \hat{\mathbf{e}}_3$.

The exchange and dipolar contributions give the magnetic energy $E$. The exchange energy is determined by $E_{\text{ex}} = \mathcal{S}A\int\mathcal{E}_{\text{ex}}\,ds$, where $A$ is the exchange stiffness, $\mathcal{S}$ is the cross-section area, and the integral is performed along the NW length. The exchange energy density is given by \cite{Sheka-JPA} $\mathcal{E}_{\text{ex}} = \mathcal{A}^2+\mathcal{B}^2$, where  $\mathcal{A}=\left(\Omega'+\kappa\cos\Phi\right)$ and $\mathcal{B}= \left(\Phi'\,\sin\Omega-\kappa\cos\Omega\sin\Phi\right)$. The dipolar energy, $E_\text{d}$, is determined by assuming that the DW lies inside an ellipsoid. This approximation is valid to describe magnetic properties of thin, narrow, and curved stripes hosting inhomogeneous magnetization textures \cite{Sheka-JPA}, including DWs \cite{Mougin-EPL,Bittencourt-APL,Yershov-Auto,Yershov-variable}. In this approach, $E_{\text{d}}=\mathcal{S}\int\mathcal{E}_{\text{d}}\,ds$, where $\mathcal{E}_{\text{d}} = 2\pi M_s^2\sin^2\Omega \left(N_2\cos^2\Phi + N_3\sin^2\Phi\right)$ and $M_s$ is the saturation magnetization. Here, $N_2$ and $N_3$ are the demagnetizing factors associated with $\hat{\mathbf{e}}_2$ and $\hat{\mathbf{e}}_3$ directions, respectively.

The magnetization dynamics is described by the Landau-Lifshitz-Gilbert-Slonczewski equation (LLGS)  \cite{LL,Gilb,Zhang-Li} as follows

\begin{equation}\label{LLG}
 \begin{array}{cl}
		\dot{\mathbf{m}} = \dfrac{\gamma}{M_s}\mathbf{m}\times\dfrac{\delta E}{\delta\mathbf{m}} + \alpha\mathbf{m}\times\dot{\mathbf{m}} + \mathbf{\Gamma}_u \, ,
  \end{array}
\end{equation}

\noindent where $\mathbf{\Gamma}_u = \mathbf{m}\times\left[\mathbf{m}\times\left(\mathbf{u}\cdot\mathbf{\nabla}\right)\mathbf{m}\right] + \beta\mathbf{m}\times\left(\mathbf{u}\cdot\mathbf{\nabla}\right)\mathbf{m}$, $\gamma$ is the gyromagnetic ratio, and $\alpha$ is the damping parameter. Here, $\mathbf{u}=-\mu_B \rho\, \mathbf{j}/[e M_s (1+\beta^2)]$, $\mathbf{j}$ is the spin-polarized current density, $\rho$ is the current polarization ratio, $\mu_B$ is the Bohr magneton, $e$ is  the  electronic  charge, and $\beta$ gives the degree of non-adiabaticity. 

An alternative to analyze the magnetization dynamics is using the Lagrangian formalism \cite{Thiaville} 

\begin{equation}\label{euler-lagrange}
    \frac{\delta \mathcal{L}}{\delta \zeta} - \frac{d}{dt}\left( \frac{\delta \mathcal{L}}{\delta \dot{\zeta}}\right) = \frac{\delta \mathcal{F}}{\delta\dot{\zeta}} \, \,  ,  \zeta \, \in \{\Omega, \Phi\} \, ,
\end{equation}

\noindent where $\delta/\delta \zeta$ and $\delta/\delta \dot{\zeta}$ are variational derivatives. The Lagrangian, $\mathcal{L}$, and Rayleigh dissipation function, $\mathcal{F}$, are given by

\begin{subequations}
    \begin{equation}\label{Lagrangeana-paramet.(1)}
  \displaystyle      \mathcal{L} = -\dfrac{M_s\mathcal{S}}{\gamma}\int \Phi\,\sin\Omega\left(\dot{\Omega}+u\Omega'\right)ds - E - E^{(u)} \, , \\
\end{equation}
\begin{eqnarray}\label{Dissipation-function-paramet(1)}
  \displaystyle   \mathcal{F} = \dfrac{M_s\mathcal{S}}{\gamma}\int \Bigg[\dfrac{\alpha}{2}\left[\dot{\Omega}^2+(\dot{\Phi}\,\sin\Omega)^2\right] \nonumber\\
  + u\beta \left(\mathcal{A}\,\dot{\Omega}+\mathcal{B}\,\dot{\Phi}\,\sin\Omega\right)\Bigg]ds \, ,
\end{eqnarray}
\end{subequations}

\noindent where $E^{(u)} = {M_s u \,\mathcal{S}}{\gamma}^{-1}\int \kappa\sin\Omega\sin\Phi\, ds$. We highlight Eq. (\ref{euler-lagrange}) is equivalent to Eq. \eqref{LLG} \cite{Yershov-helice, Thiaville}. 

Using the Lagrangian formalism, we describe a DW by a collective variables approach \cite{Thiaville,Slonk}, in which it is represented by the ansatz $\Omega(s, t) = 2\arctan\left\{\exp[p(s-q(t))/\lambda]\right\}\equiv\Omega$ and $\Phi = \phi(t)\equiv\phi$, where $p=+ 1\,(-1)$ denotes the head-to-head (tail-to-tail) DW type, $q$ is the DW  position, and $\lambda$ is the DW width, which is considered constant during the DW propagation. This approach is valid for bent NWs where $\kappa\lambda\lesssim1$ \cite{Yershov-PRB,Yershov-helice}. In this case, the exchange and dipolar energies can be evaluated as

\begin{subequations}
\begin{equation}\label{Ex-transverse-DW}
    E_{\text{ex}} = A\mathcal{S}\left(\frac{2}{\lambda} + 2 \mathcal{I}_\text{ex}^{(1)}\cos\phi - \mathcal{I}_\text{ex}^{(2)}\sin^2\phi\right) \, ,
\end{equation}
\begin{equation}\label{Ed-transverse-DW}
    E_{\text{d}} \approx 4\pi\lambda M_s^2\mathcal{S}\left(N_2\cos^2\phi + N_3\sin^2\phi\right) \, ,
\end{equation}
\end{subequations}

\noindent where $\mathcal{I}_\text{ex}^{(1)} = \int\Omega'\,\kappa ds$ and $\mathcal{I}_\text{ex}^{(2)} = \int\kappa^2 \sin^2\Omega\, ds$. We highlight that in the limit $\kappa\lambda \ll 1$ these integrals give  \cite{Yershov-PRB} $\mathcal{I}_\text{ex}^{(1)} \approx p \pi \kappa_q$ and $\mathcal{I}_\text{ex}^{(2)} \approx 2\lambda\kappa_q^2$, where $\kappa_q$ is the curvature at the DW position. The terms that are proportional to $\mathcal{I}_\text{ex}^{(1)}$ and $\mathcal{I}_\text{ex}^{(2)}$ correspond to the curvature-induced effective Dzyaloshinskii–Moriya and the anisotropy interactions \cite{Sheka-JPA,Yershov-PRB,Yershov-helice}, respectively, which, together with the dipolar interaction, determine the magnetization properties in curved NWs \cite{Bittencourt-APL,Yershov-Auto,Yershov-variable}. The collective variables approach allows us to obtain the integrals in $\mathcal{L}$ and $\mathcal{F}$, yielding the effective functions

\begin{subequations}
\label{LagEff}    
\begin{equation}\label{LagrangeanaEff-paramet.(1)}
            \mathcal{L}_{\text{eff}} = -\dfrac{2M_s\mathcal{S}}{\gamma}\left[p\,\phi(u-\dot{q}) + \dfrac{u}{2}\,\mathcal{I}_\text{u}^{(1)}\sin\phi\right] - E\, ,
\end{equation}
\begin{equation}\label{DissipationEff-paramet.(1)}
    \begin{array}{cl}
         \mathcal{F}_{\text{eff}} = \dfrac{M_s\mathcal{S}}{\gamma}\Bigg[\alpha\left(\lambda\,\dot{\phi}^2+\dfrac{\dot{q}^2}{\lambda}\right) 
         - u\beta\bigg(\dfrac{2}{\lambda}\,\dot{q} \\
         \\ +\dfrac{p\dot{q}}{\lambda}\,\mathcal{I}_\text{u}^{(1)}\cos\phi +\dfrac{\dot{\phi}}{2}\,\mathcal{I}_\text{u}^{(2)}\,\sin\phi\bigg) \Bigg] \,,          
    \end{array}
\end{equation}
\end{subequations}

\noindent where $\mathcal{I}_\text{u}^{(1)} = \int \kappa\,\sin \Omega\, ds$ and $\mathcal{I}_\text{u}^{(2)} = \int \kappa\,\sin 2\Omega\, ds = -2p\lambda\int\kappa'\sin\Omega\, ds$ are curvature-dependent spin-transfer torque parameters. In the limit $\kappa \lambda \ll 1$, we obtain $\mathcal{I}_\text{u}^{(1)} \approx \pi\lambda\kappa_q$. For NWs with a constant curvature,  we obtain $\mathcal{I}_\text{u}^{(2)} = 0$, recovering the results from Yershov \textit{et al.} \cite{Yershov-helice}.  Another useful property of these integrals is $\mathcal{I}_\text{u}^{(2)} = -2p\lambda\,\partial \mathcal{I}_\text{u}^{(1)}/\partial q$. Although the collective variable approach was first developed for straight systems, it describes the DW dynamics in bent NWs without torsion and small curvature \cite{Yershov-PRB,Yershov-helice}. The dynamical equations regard the pair of time-dependent coordinates $\{q(t), \phi(t)\}$ and can be obtained by inserting Eqs. \eqref{LagEff} into \eqref{euler-lagrange}, leading to

\begin{subequations}\label{PosPhase}
\begin{equation}
    \dot{q} = -\dfrac{p\gamma}{2M_s S}\mathcal{Q}_\phi + \frac{up}{2}\left(\mathcal{Y}_{u} - {\beta\,\mathcal{Q}_{\text{eff}}^{(2)}}\right) + p\alpha\lambda\dot{\phi} \, ,
    \label{qdot}
\end{equation}

\begin{equation}
    \dot{\phi} = \dfrac{p\gamma}{2M_s S}\mathcal{Q}_q+\frac{u}{2\lambda}\left(\beta\,\mathcal{Y}_u+ {\mathcal{Q}_{\text{eff}}^{(2)}}\right) - \dfrac{p\alpha}{\lambda}\dot{q} \, .
    \label{phidot-2}
\end{equation}

\end{subequations}

\noindent where $\mathcal{Y}_{u}=2p + \mathcal{Q}_\text{eff}^{(1)}$ and $\mathcal{Q}_\zeta \equiv - \partial _\zeta E$. Here, $\mathcal{Q}_\text{eff}^{(1)}$ and $\mathcal{Q}_{\text{eff}}^{(2)}$ are current-driven terms associated with the NW geometry, given by  $\mathcal{Q}_\text{eff}^{(1)}={\mathcal{I}_\text{u}^{(1)}}\cos\phi$ and $
 \mathcal{Q}_{\text{eff}}^{(2)} = ({\mathcal{I}_{\text{u}}^{(2)}}/2)\sin\phi\, .$
Eqs. \eqref{PosPhase} together with the term proportional to $\mathcal{Q}_{\text{eff}}^{(2)}$ are the core of our work. Firstly, one can notice that for $\kappa'(s)=0$, $\mathcal{Q}_{\text{eff}}^{(2)}=0$. Nevertheless, if $\kappa'(s) \neq 0$, our results evidence the emergence of a position and phase-dependent CSTT term, $\mathcal{Q}_{\text{eff}}^{(2)}$, which vanishes when $\phi=0$ or $\phi=\pi$, and whose extreme values occur for $\phi=\pm\pi/2$. An interesting consequence of the CSTT appears if the DW moves under a steady regime ($\dot{\phi}=0$). In this case, unlike what occurs in a NW with constant curvature \cite{Thiavile-EPL,Yershov-helice}, the DW velocity does not vanish even when $\beta=0$, being evaluated as $ \dot{q} ={\lambda\gamma}({2\alpha\,M_s S})^{-1}\mathcal{Q}_q +(2\alpha)^{-1}p\,u\,\mathcal{Q}_{\text{eff}}^{(2)}$. This result evidences that the curvature gradient yields two effective forces on the DW. The first contribution comes from the exchange-driven curvature-induced effective interactions \cite{Yershov-PRB, Sheka-JPA} while the second contribution originates from the CSTT.  

The CSTT can be interpreted by writing $\mathbf{\Gamma}_u = -\gamma \,\mathbf{m}\times \mathbf{H}_u$, where $\mathbf{H}_u$ is the effective field associated with the current density (see details in the Supplemental Materials). To simplify our analysis, we focus only on the $\mathbf{H}_u$ curvature-dependent component tangent to the NW that does not vanish for $\phi=\pm\pi/2$,

\begin{equation}
   \mathbf{H}_{t} = \frac{u\kappa}{2\gamma}\sin2\Omega\sin\phi\,\hat{\mathbf{e}}_1\,,
   \label{tangential-field}
\end{equation}

\noindent from which we obtain $\mathcal{Q}_{\text{eff}}^{(2)} = \gamma\,u^{-1}\, \int  (\mathbf{H}_{t}\cdot\hat{\mathbf{e}}_1)\,ds$. Therefore, the CSTT results from the different strengths of the interaction between the conduction electrons and the local magnetization. The net effect of this local interaction is a position and phase-dependent effective field acting on the DW. Consequently, for variable $\kappa$, the DW translational and rotational motion can be reinforced or blocked depending on $\phi$. If $\kappa$ is constant, $\mathbf{H}_{t}$ has the same modulus and different directions when evaluated in opposite positions with respect to the DW center, resulting in a vanishing effect on the DW motion. 

\begin{figure}[!h]
    \centering
    \includegraphics[width=7.5cm ,angle=0]{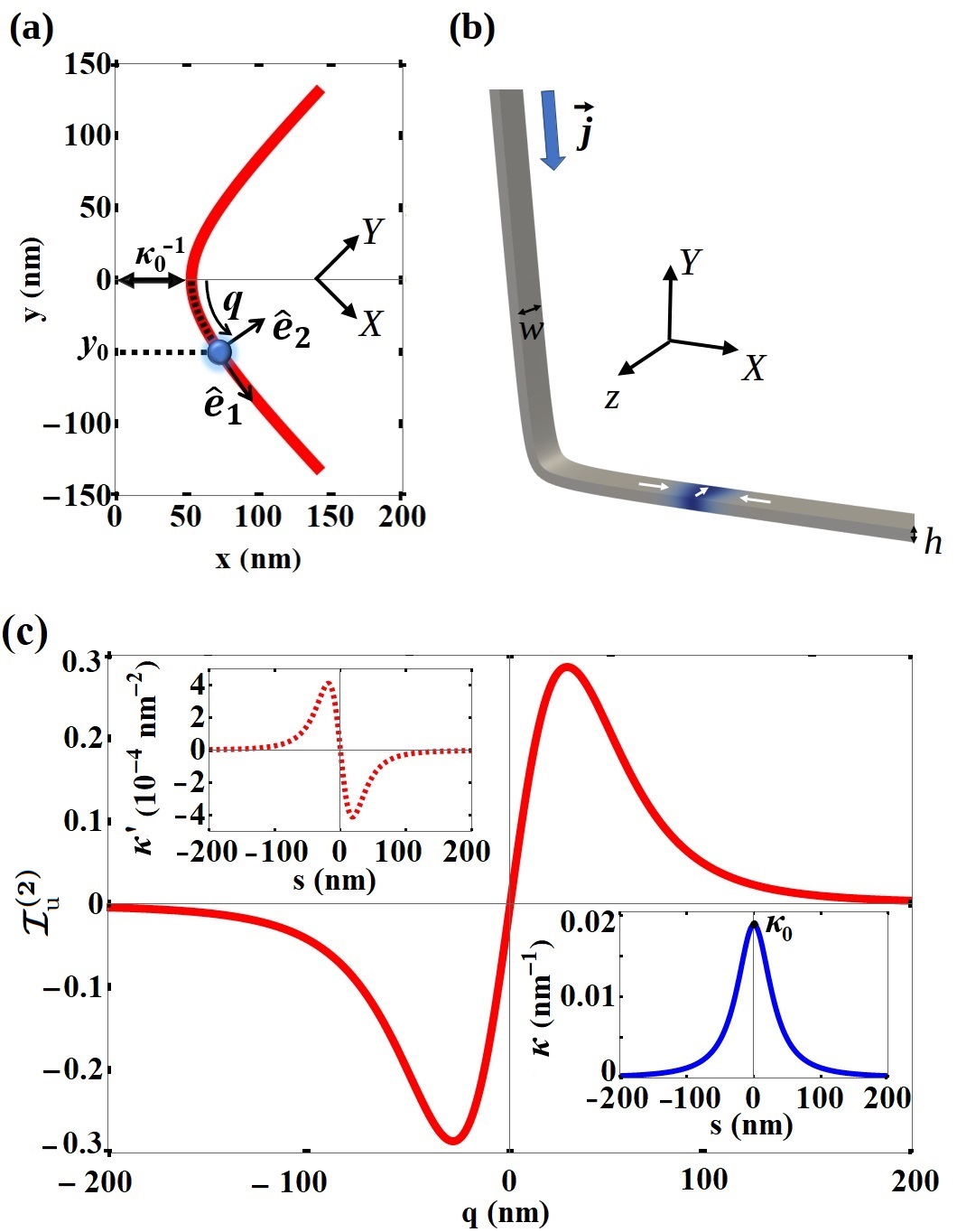}
    \caption{\textbf{(a)} NW geometry for $\kappa_0 = 6\pi \times 10^{-3}$ nm$^{-1}$. The $XY$-axis is a rotated frame that relates the DW position (blue dot) and the projection along the $y$-axis. \textbf{(b)} depicts the structure of the simulated system with  a head-to-head DW shown in blue. Fig. \textbf{(c)} shows the behavior of ${\mathcal{I}_{\text{u}}^{(2)}}$ as a function of the DW position. The insets illustrate the NW curvature (blue line) and the curvature gradient (red-dotted line).}
    \label{fig.1}
\end{figure}

\begin{figure}[!h]
    \centering
    \includegraphics[width=8.25cm ,angle=0]{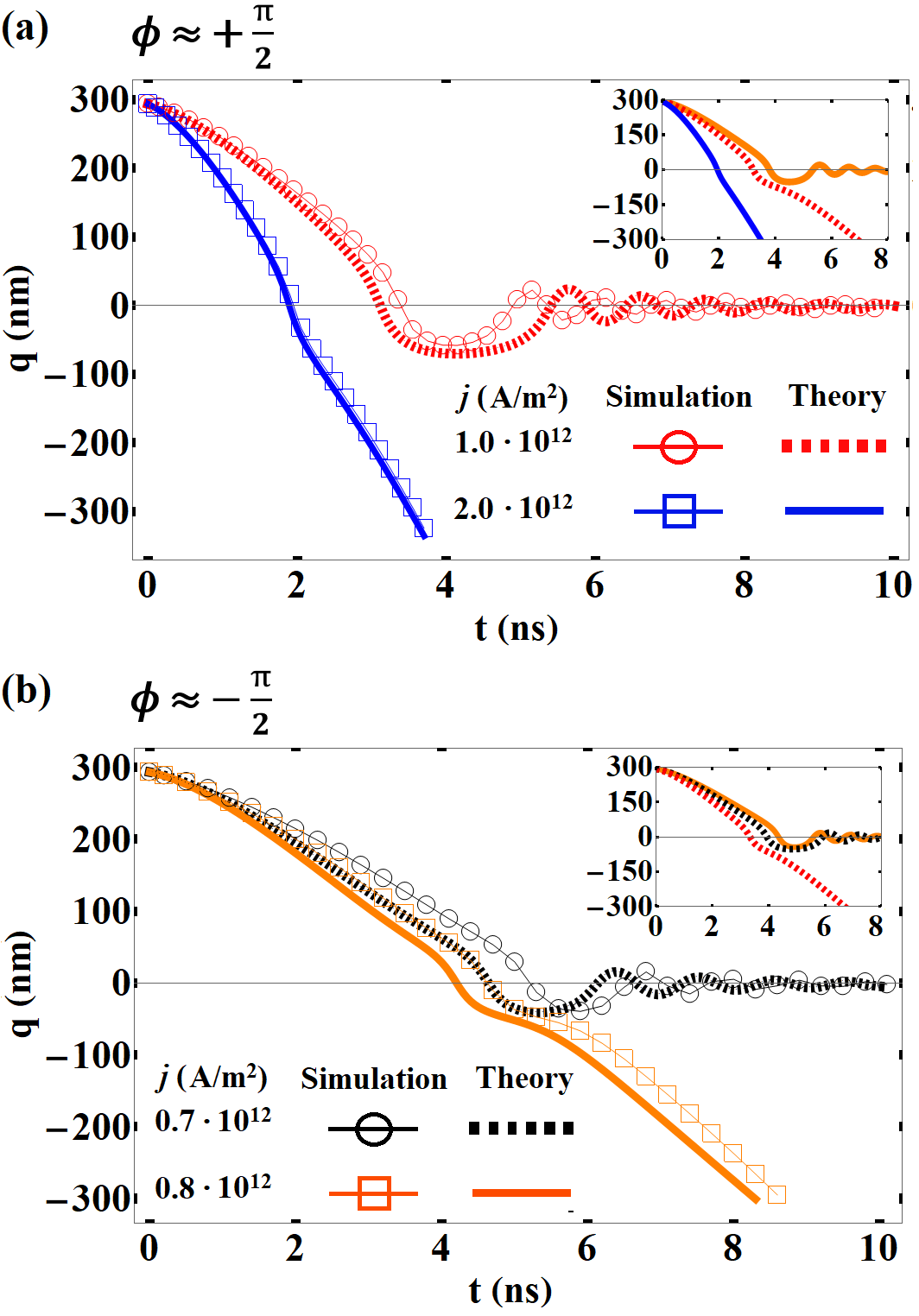}
    \caption{DW position as a function of time for different electric current values. (a) and (b) show the results of a DW pointing along the $+\hat{\mathbf{z}}$ and $-\hat{\mathbf{z}}$ direction, respectively. The insets illustrate  the DW position if the CSTT did not exist.}
    \label{fig.2}
\end{figure}

As a particular case, we can use the above-described formalism to analyze the CSTT effects on a DW displacing along a  NW curved as a hyperbole. We consider that the hyperbole lies on the $xy$-plane, parametrized as $\boldsymbol{\chi}(y) = [{y^2 + \kappa_0^{-2}}]^{1/2}\, \hat{\mathbf{x}} + y \,\hat{\mathbf{y}} $. In this case, the NW curvature is $\kappa = \kappa_0/\left(1+2y^2\kappa_0^2\right)^{3/2}$, where $\kappa_0$ refers to the maximum value of the curvature, occurring at $y=0$, as shown in Fig. \eqref{fig.1}(a). The natural parameter related with this parametrization is $ds = - h_y dy$, where $h_y = [{1 + y^2/(y^2+\kappa_0^{-2})}]^{1/2}$.
An arbitrary position on the NW is defined by an elliptic integral of the second kind, $q = -\int_0^{y_0} h_y dy$, which associates the coordinate $y_0$ to the projection of $q$ onto $y$-axis. Here, we present the results for a NW with $\kappa_0 = 6\pi \times 10^{-3}$ nm$^{-1}$, $w=20$ nm, and $h=5$ nm. We have also obtained the DW dynamics for different values of $\kappa_0$, presented in the Supplemental Materials. The used  parameters are $A=1.3\times10^{-11}$ J/m and $M_s=8.6\times10^5$ A/m, a damping parameter $\alpha = 0.01$, a non-adiabaticity degree $\beta = 0.04$, and a polarization ratio $\rho=0.4$. We adopt $\lambda\approx14.5$ nm as the averaged value of the DW width obtained using the micromagnetic simulations presented below. Finally, our analysis considers the specific case of $p=1$ and a range of currents in which the DW moves without rotating around the NW axis.

We start numerically solving Eqs. \eqref{PosPhase} for different values of the electric current. Results evidence phase-dependent DW dynamics with two regimes depending on the electric current, as shown in Fig. \ref{fig.2}. Indeed, for $j\lesssim1\times10^{12}$ A/m$^2$ (red-dashed line) and  $\phi\approx\pi/2$, the DW displaces through the NW and oscillates around the maximum curvature point defined by $q=0$, until stopping. For $j\gtrsim2\times10^{12}$ A/m$^2$ (blue line), the DW crosses the region with maximum curvature with a velocity that increases with $j$. When the DW phase is $\phi\approx-\pi/2$, a similar phenomenology occurs, but the threshold current separating both regimes is smaller.  Indeed, Fig. \ref{fig.2}(b) reveals that the DW continues traveling through the maximum curvature point even when $j\approx8\times10^{11}$ A/m$^2$ (orange line), being pinned for $j\approx7\times10^{11}$ A/m$^2$ (black-dashed line). We also determined the DW dynamics for other current density values, but no qualitative changes were observed.

To clarify the interplay between the DW phase and the CSTT,  we  calculated ${\mathcal{I}_{\text{u}}^{(2)}}$ as a function of $q$ (see Fig. \ref{fig.1}(c)). The proportionality between ${\mathcal{I}_{\text{u}}^{(2)}}$ and $\kappa'(s)$ is evident, with the maximum magnitude of $\mathcal{Q}_{\text{eff}}^{(2)}$ occuring at the point of greater curvature gradient, being in the order of $\mathcal{Q}_{\text{eff}}^{(2)}\approx0.15$, while $\beta\mathcal{Y}_{\text{u}}\approx0.08$. To deeply understand the DW pinning, we performed numerical calculations by assuming that $\mathcal{Q}_{\text{eff}}^{(2)}=0$. Results are presented in the insets of Figs.\ref{fig.2}(a) and (b), showing that if the CSTT term disappears, the DW dynamics would be phase-independent, allowing us to conclude that the phase-dependence of the threshold current value is due to the CSTT. In conclusion, for $\phi = \pi/2$, the main pinning effect comes from the CSTT. In contrast, for $\phi = -\pi/2$, the CSTT term pushes the DW, contributing to its propagation through the NW.

To corroborate our results, we have performed micromagnetic simulations using the Nmag code \cite{Fangohr}. The simulated system consists of a nanostripe with length $1200$ nm, meshed using GMSH \cite{Gmsh} with a characteristic length of $1.5$ nm. This cell size ensures that the distance between nodes for any element is smaller than the exchange length. The current density direction was obtained using COMSOL Multiphysics \cite{Comsol} using the AC/DC module. In all simulations, the nanostrip was initially magnetized in a head-to-head configuration by considering the initial position of the domain wall at $300$ nm from the maximum curvature point. To set the direction of the domain wall, a small region at the center of the wall was magnetized in the $\pm \hat{z}$ direction. The system was then relaxed  for 5 ns. The position of the center of the domain wall was then calculated using the arithmetic mean of the position of all points satisfying the condition $m_z>0.95$ when $\phi = +\pi/2$ and $m_z<-0.95$ when $\phi = -\pi/2$. The visualization of the magnetization configuration was done using Paraview \cite{Paraview}. After relaxing the system, we apply different electric current values to analyze the DW motion. The DW position obtained from micromagnetic simulations is presented by circles and squares in Fig. \ref{fig.2}, showing an excellent agreement with the results obtained with the analytical model. The small quantitative differences between both results are related to deformations on the DW shape, not considered in the theoretical model. Therefore, the rigid DW model is an excellent approach to describe the DW dynamics in the analyzed cases. 

\begin{figure}[!h]
    \centering
    \includegraphics[width=8.cm ,angle=0]{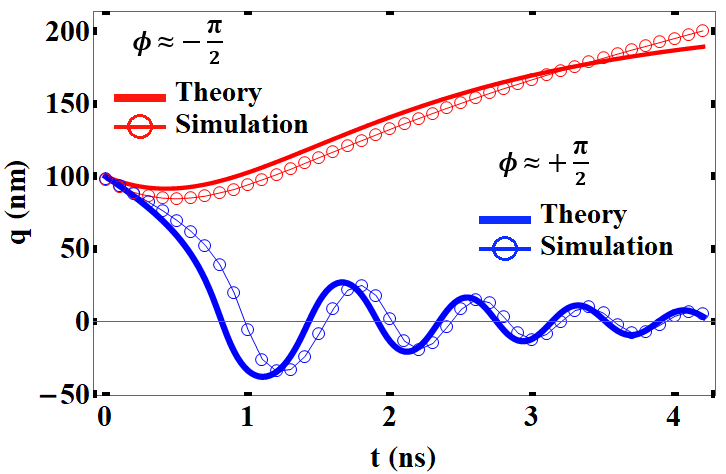}
    \caption{DW position as a function of time for $\beta=0$ and $j=2\times 10^{12}$ A/m$^2$. Blue and red curves depict the results for $\phi\approx\pi/2$ and $\phi\approx-\pi/2$, respectively.}
    \label{fig.3}
\end{figure}

The chiral effect becomes more evident when we analyze the DW motion for $\beta=0$. Results are illustrated in Fig. \ref{fig.3}, for a current density $j=2\times10^{12}$ A/m$^2$. Lines and symbols show the results from the theoretical model and micromagnetic simulations. For $\phi\approx\pi/2$ (blue), the DW displaces along the NW to the region of maximum curvature, where it is finally pinned. However, for $\phi\approx-\pi/2$ (red), the CSTT has a negative sign and forces the DW to propagate in the opposite direction.  This DW backward motion for $\beta=0$ suggests the existence of a threshold value, $\beta_c$, at which the DW velocity vanishes. Considering that the DW is far from $y=0$ and $\phi\approx-\pi/2$, it does not move when it arrives in a position where the relation $\beta_c\approx\mathcal{I}^{(2)}_\text{u}/{4}$ is satisfied. For the analyzed system, $\beta_c \approx 0.01$ and $0.002$ for $q \approx 100$ nm and $150$ nm, respectively. For a straight NW, $\beta_c=0$, in agreement with results in Ref. \cite{Thiaville}.

At this stage, we would like to highlight that the CSTT is not present only in current-driven DW motions, but it is a more generalized effect that appear in other current-induced spin propagation processes.  For example, in spin wave propagation along curved nanowires.  Indeed, by considering $\Omega=\pi/2+\vartheta$ and $\phi=\pm\pi/2+\varphi$, where $\vartheta\equiv\vartheta(s,t)$ and $\varphi\equiv\varphi(s,t)$ are small perturbations, we obtain a chiral curvature-dependent effective field component tangent to the NW, defined by Eq. \eqref{tangential-field}, and given by $\mathbf{H}_t=\mp u\,\kappa\,\vartheta\,\gamma^{-1}\,\hat{\mathbf{e}}_1$, where the sign $\mp$ is associated with $\phi\approx\pm\pi/2$. 

In summary, using a theoretical approach and simulations we have predicted the existence of a chiral spin-transfer torque phenomenon present when a DW propagates along magnetic systems with a curvature gradient. The origin of the CSTT relates to the emergence of an effective field, tangent to the NW. We have determined the DW dynamics in a hyperbolic NW, showing a phase dependent DW pinning and unpinning. Therefore, the CSTT affects the threshold value of the electric current needed for  DW propagation with no pinning in the curved region. Our results evidence that the CSTT is a more general phenomenon, which can also appear for other magnetization collective modes propagating in nanosystems with curvature gradient. Also, the CSTT could be behind experimental asymmetries in the dynamical behavior of head-to-head and tail-to-tail DWs during their current-driven motion in curved NWs \cite{Parkin-Nat3,Meier-JAP,Hu-SciRep}. The CSTT should be considered in applications demanding the spin transport in curved devices and future experimental and theoretical works, since it presents opportunities for designing current-controlled spintronic devices.

\paragraph{Acknowledgements} In Brazil, we thank the agencies Capes (Finance Code 001), CNPq (Grant No. 305256/2022), and Fapemig (Grant No. APQ-00648-22). Funding is acknowledged from Fondecyt Regular 1200867, 1220215, 1230515 and Financiamiento Basal para  Centros  Cient\'ificos  y  Tecnol\'ogicos  de  Excelencia AFB220001. Powered@NLHPC: This research was partially supported by the supercomputing infrastructure of the NLHPC (ECM-02).

\end{document}